\documentclass[graybox]{svmult}

\usepackage{amssymb,amscd,amsmath,amsthm}

\usepackage{type1cm}        
\usepackage{makeidx}         
\usepackage{graphicx}        
\usepackage{multicol}        
\usepackage[bottom]{footmisc}

\usepackage{newtxtext}       %

\makeindex             

\begin{document}

\title*{$P$-adic dynamical systems for $p$-adic $\lambda$-models on the Cayley
trees}
\author{Farrukh Mukhamedov\orcidID{0000-0001-5728-6394} and\\ Otabek Khakimov\orcidID{0000-0002-8918-9094}}
\institute{Farrukh Mukhamedov \at Department of Mathematical Sciences,
College of Science, The United Arab Emirates University, Al Ain, UAE, \email{far75m@gmail.com,\ farrukh.m@uaeu.ac.ae}
\and Otabek Khakimov \at Department of Algebra and its applications,\\
V.I.Romanovsky Institute of Mathematics, Tashkent, Uzbekistan, \email{khakimovo86@gmail.com}}
%
%
\maketitle


\abstract{We review and propose to use of associated dynamical system
to explore the phase transition phenomena in $p$-adic  statistical mechanics setting, by means of the renormalization
techniques. Main focus of the paper is the $p$-adic
$\lambda$-model on the Cayley tree. We study generalized $p$-adic quasi Gibbs measures for the
$\lambda$-model. These measures associate recurrence equations that determine dynamical systems for
the model. A main point is to verify and confirm that the indicated
predictions from a dynamical system point of view are indeed true.}

\section{Introduction}
\label{sec:1}
Since Wilson's pioneering contributions in the early 1970s \cite{Wil} -- building
on the foundational insights of Kadanoff, Widom \cite{WK82}. Fisher \cite{Fish},
and others in the preceding decade -- the renormalization group (RG) has profoundly
shaped modern statistical physics \cite{FTC}. Beyond offering a powerful analytical
toolkit for characterising and quantifying both static and dynamic critical
phenomena near continuous phase transitions, where strong interactions,
fluctuations, and correlations dominate, the RG has furnished a unifying
conceptual framework and mathematical language that now permeates the
theoretical treatment of diverse, intricate many-body systems \cite{Ta}.

The RG framework has been extensively deployed in statistical mechanics
and has generated a wealth of important results. Notably, studies of
spin-model phase transitions on hierarchical lattices reveal that such
geometries often admit exact evaluations of diverse thermodynamic
quantities \cite{Bax,G}. One of the simplest hierarchical structures
is the Cayley (or Bethe) tree \cite{Ost}. Although this lattice is an
idealised graph rather than a realistic crystalline substrate, spin
systems defined on it allow rigorous, closed-form calculations that
remain analytically intractable on regular lattices \cite{Roz}.
It is generally conjectured that many of the tree's striking thermal
properties survive, at least qualitatively, in more realistic settings.

The RG method is frequently illustrated through the Ising model,
owing to its broad theoretical significance and numerous applications
\cite{Bax,Egg}. Clearly, RG transformation basically depends on the construction
 of hierarchical
lattice and model. The simplest one is governed by rational
functions (see \cite{ARS,B0,B1,DS11,DKM,KM2,MR3,RS}). Recently, \cite{DSI}
has provided an important and interesting
relation between the phase transition and chaotic behavior of the RG
transformation (see, also \cite{MFKh16,MK17}).
We notice that in \cite{ALS17,JS24,M13,M15,MA1,MK18,MK201} the RG method has
been applied to the detection of the phase
transitions for $p$-adic Potts models on Cayley trees.
In this way, it is noted that chaotic $p$-adic dynamical systems have
immense applications in coding theory \cite{FL2,TVW,Wo}. We refer the reader
for more information about $p$-adic dynamical systems to \cite{AKh,KhN}.

A natural generalization of the Ising model is the so-called $\lambda$-model
on the Cayley tree \cite{M04,R1}, whose rich structure captures virtually
every nuance of statistical-mechanical behaviour. Central to the theory of
Gibbsian lattice systems is the characterisation of the infinite-volume (limiting)
Gibbs measures associated with a given Hamiltonian \cite{G,Roz}.

Conversely, a substantial body of work argues that $p$-adic numbers may
offer a more faithful and precise mathematical framework for modelling
microworld phenomena (see, e.g., \cite{Kh1,Man,VVZ,V1}). This view is intimately
linked to longstanding questions concerning the structure of spacetime at the
Planck length, $\lambda_{P} = (G\hbar/c^{3})^{1/2} \approx 1.6\times 10^{-33}\,\mathrm{cm}$,
a scale whose physical nature has stimulated vigorous debate in the literature \cite{DFR,Man,Sn}.
Since the 1980s, this perspective has motivated the development of diverse models
formulated within the machinery of $p$-adic analysis \cite{ADFV,FO,MP}.
Much of this research focuses on quantum-mechanical systems expressed through
the equations of mathematical physics \cite{AKS,ADV,AZ,ABK,Kh1,Kh2,V2,VVZ,ZT}.
For a comprehensive overview of recent advances, the reader is referred
to \cite{DKKV,Drag2}.

One of the earliest incursions of $p$-adic analysis into quantum physics
arose within the framework of quantum logic \cite{BC}.  The resulting model
is of particular interest because it resists description in terms of ordinary
real-valued probabilities \cite{Kh2,Ko,MP,VVZ}.  This limitation prompted the
development of specifically $p$-adic probability models \cite{kas1,kas2,kas3,K3,KhN,KYR,Y}.
Building on subsequent advances in $p$-adic measure theory \cite{Kh07,KL,Lu}, researchers
formulated comprehensive theories of $p$-adic and, more broadly,
non-Archimedean-stochastic processes.
These theories permit the construction of extensive families of stochastic
processes directly from their finite-dimensional distributions \cite{GMR}.
Within this probabilistic framework, a programme of $p$-adic statistical
mechanics has taken shape \cite{ALS17,GRR,GMR1,KhakSMJ,KMM,Mq,MDA,MR1,MR2,RKh,RT,SA15}.
In particular, $p$-adic Ising and Potts models with nearest-neighbour
interactions have been rigorously analysed on Cayley trees (see \cite{MK21c}
for a recent review on the Potts models).  Beyond these lattice systems,
further $p$-adic models have been proposed for describing complex hierarchical
structures \cite{KK2,KK3}.

This paper examines phase transitions in $p$-adic statistical models
through renormalization methods formulated in a measure-theoretic setting.
Because renormalization naturally entails analysing the dynamical system
generated by the model, our approach draws on techniques from both $p$-adic
dynamical systems and $p$-adic probability theory. The past decade has
witnessed rapid progress in $p$-adic dynamics \cite{FL4,KhN,Sil1}, building
on the seminal non-Archimedean work of Herman and Yoccoz \cite{HY}. More
recently, research activity has intensified across $p$-adic and broader
algebraic dynamical systems (see, e.g., \cite{AKh,AV1,AV2,DKM,FL2,Kh22,L,MM1,MR4,S24,TVW,Wo});
an exhaustive, regularly updated bibliography is curated by Silverman \cite{Sil2}.

This study advances the analysis of the $p$-adic $\lambda$-model initiated
in \cite{KM,KMR}, where attention was confined to proving the uniqueness of the
corresponding $p$-adic Gibbs measures. We notice that the classical $\lambda$-models have been investigated in several papers \cite{M22,MJP2018,R1}.
Recent work \cite{M12,M13} has refined
the notion of criticality by introducing two distinct concepts -- `phase transition'
and `quasi-phase transition'. In the classical (real-valued) setting,
dynamical-systems methods have substantially deepened our understanding
of such critical behaviour, and their interplay with chaos theory has
generated innovative conceptual frameworks in diverse physical contexts \cite{E}.
Motivated by these developments, the present paper applies a renormalisation-group
approach to demonstrate and characterise phase transitions in the $p$-adic regime.

Renormalization techniques constitute one of the most powerful tools for
analysing phase transitions in theoretical and mathematical physics. In the
real-valued setting, these methods have markedly advanced our understanding
of complex model behaviour, while their synergy with chaos theory has inspired
new conceptual frameworks across disparate physical systems \cite{E}. Motivated
by these successes, the present study adopts a renormalization-group perspective
to examine critical phenomena in the $p$-adic context. Specifically, we investigate
the $p$-adic $\lambda$-model on the Cayley tree; taking $\lambda(x,y)=Nxy$
recovers the $p$-adic Ising model previously analysed in \cite{M13,MD15}.

The paper is structured as follows. Section~2 introduces preliminary definitions and auxiliary results concerning the Cayley tree and \( p \)-adic analysis. In Section~3, we summarize essential background on the \( \lambda \)-model on the Cayley tree, following the developments in~\cite{KMR,M04,M15,M22}. Section~4 focuses on a specific case of the interaction function \( \lambda \), which gives rise to the \( p \)-adic Ising model. Here, we review key findings from~\cite{MAD17,MD15,MDA,MK17,MK21} and establish connections with results on periodic Gibbs measures for the Ising model discussed in~\cite{RA241}. We also formulate several open problems related to the chaotic dynamics of the Ising mapping and their implications for the proliferation of weakly periodic Gibbs measures. Finally, Section~5 revisits results from~\cite{KM,M15,MA15,MD15}, which provide the foundation for posing open questions regarding the chaotic behavior of the associated dynamical system for the \( \lambda \)-model.

\section{Preliminaries}
\label{sec:2}
\subsection{$p$-adic numbers}
In what follows $p$ will be a fixed prime number. The set $\mathbb Q_p$
is defined as a completion of the rational numbers $\mathbb Q$ with
respect to the norm $|\cdot|_p:\mathbb Q\to\mathbb R$ given by
\begin{eqnarray}
|x|_p=\left\{
\begin{array}{ll}
  p^{-r}, & \mbox{if}\ x\neq 0,\\
  0, & \mbox{if}\ x=0,
\end{array}
\right.
\end{eqnarray}
here, $x=p^r\frac{m}{n}$ with $r,m\in\mathbb Z$, $n\in\mathbb N$,
$(m,p)=(n,p)=1$. The absolute value $|\cdot|_p$ is
non-Archimedean, meaning that it satisfies the \textit{strong
triangle inequality} $|x + y|_p \leq \max\{|x|_p, |y|_p\}$. We
recall a nice property of the norm, i.e. if $|x|_p>|y|_p$ then
$|x+y|_p=|x|_p$. Note that this is a crucial property which is
proper to the non-Archimedenity of the norm.

Any $p$-adic number $x\in\mathbb Q_p$, $x\neq 0$ can be uniquely
represented in the form
\begin{equation}\label{canonic}
x=p^{\gamma(x)}(x_0+x_1p+x_2p^2+...),
\end{equation}
where $\gamma=\gamma(x)\in\mathbb Z$ and $x_j$ are integers, $0\leq x_j\leq
p-1$, $x_0>0$, $j=0,1,2,\dots$ In this case $|x|_p=p^{-\gamma(x)}$.

For each $a\in \mathbb Q_p$, $r>0$ we denote $$ B(a,r)=\{x\in \mathbb Q_p :
|x-a|_p< r\}, \ \ \mathbb Z_{p}=\left\{ x\in \mathbb Q_p:\
|x|_{p}\leq1\right\}.$$

Recall that the $p$-adic exponential is defined by
$$
\exp_p(x)=\sum_{n=1}^{\infty}\frac{x^n}{n!},
$$
which converges for every $x\in B(0,p^{-1/(p-1)})$. It is known
\cite{Ko} that for any $x\in B(0,p^{-1/(p-1)})$ one has
$$ |\exp_p(x)|_p=1,\ \ \ |\exp_p(x)-1|_p=|x|_p<1. $$

Put
\begin{equation}\label{Exp}
\mathcal E_p=\{x\in\mathbb Q_p: \ |x|_p=1, \ \ |x-1|_p<p^{-1/(p-1)}\}.
\end{equation}

Note that the basics of $p$-adic analysis, $p$-adic mathematical
physics are explained in \cite{Ko,S,VVZ}.

Now, we recall some standard terminology of the theory of dynamical
systems (see for example \cite{AKh,KhN}).

Let $(f,B)$ be a dynamical system in $\mathbb Q_p$, where $f: B\ni x\to
f(x)\in B$ is some function and $B=B(a,r)$ or $\mathbb Q_p$. Denote
$x^{(n)}=f^n(x^{(0)})$, where $x^0\in B$ and
$f^n(x)=\underbrace{f\circ\dots\circ f(x)}_n$.
 If $f(x^{(0)})=x^{(0)}$ then $x^{(0)}$
is called a {\it fixed point}. Let $x^{(0)}$ be a fixed point of
an analytic function $f$. Set
$$
\lambda=\frac{d}{dx}f(x^{(0)}).
$$

The point $x^{(0)}$ is called {\it attractive} if $0\leq
|\lambda|_p<1$, {\it neutral} if $|\lambda|_p=1$, and {\it repelling} if
$|\lambda|_p>1$.

It is known \cite{KhN} that if a fixed point $x^{(0)}$ is
attractive then there exists a neighborhood $U(x^{(0)})(\subset
B)$ of $x^{(0)}$ such that for all points $y\in U(x^{(0)})$ it
holds $\lim\limits_{n\to\infty}f^{(n)}(y)=x^{(0)}$. If a fixed
point $x^{(0)}$ is  repelling, then there  exists a neighborhood
$U(x^{(0)})$ of $x^{(0)}$ such that
$|f(x)-x^{(0)}|_p>|x-x^{(0)}|_p$ for $x\in U(x^{(0)})$, $x\neq
x^{(0)}$.

\subsection{$p$-adic sub-shift}
Let $f:X\to\mathbb Q_p$ be
a mapping from a compact open set $X$ of $\mathbb Q_p$ into $\mathbb
Q_p$. We assume that (i) $f^{-1}(X)\subset X$; (ii)
$X=\bigcup\limits_{j\in I}B_{r}(a_j)$ can be written as a finite
disjoint union of balls of centers $a_j$ and of the same radius $r$
such that for each $j\in I$ (where $I$ is a finite subset of the set
of positive integers) there is an integer $\tau_j\in\mathbb Z$ such
that
\begin{equation}\label{tau}
|f(x)-f(y)|_p=p^{\tau_j}|x-y|_p,\ \ \ \ x,y\in B_r(a_j).
\end{equation}
For such a map $f$, define its Julia set by
\begin{equation}\label{J}
J_f=\bigcap_{n=0}^\infty f^{-n}(X).
\end{equation}
It is clear that $f^{-1}(J_f)=J_f$ and then $f(J_f)\subset J_f$.

Following \cite{FL2} the triple $(X,J_f,f)$ is called a $p$-adic
{\it weak repeller} if all $\tau_j$ in (\ref{tau})  are nonnegative,
but at least one is positive. We call it a $p$-adic {\it repeller}
if all $\tau_j$ in (\ref{tau}) are positive.
 For any $i\in I$, we let
$$
I_i:=\left\{j\in I: B_r(a_j)\cap
f(B_r(a_i))\neq\varnothing\right\}=\{j\in I: B_r(a_j)\subset
f(B_r(a_i))\}
$$
(the second equality holds because of the expansiveness and of the
ultrametric property). Then define a matrix $A=(a_{ij})_{I\times
I}$, called \textit{incidence matrix} as follows
$$
a_{ij}=\left\{\begin{array}{ll}
1,\ \ \mbox{if }\ j\in I_i;\\
0,\ \ \mbox{if }\ j\not\in I_i.
\end{array}
\right.
$$
If $A$ is irreducible, we say that $(X,J_f,f)$ is
\textit{transitive}. Here the irreducibility of $A$  means, for any
pair $(i,j)\in I\times I$ there is a positive integer $m$ such that
$a_{ij}^{(m)}>0$, where $a_{ij}^{(m)}$ is the entry of the matrix
$A^m$.

Given $I$ and the irreducible incidence matrix $A$ as above. Denote
$$
\Sigma_A=\{(x_k)_{k\geq 0}: \ x_k\in I,\  A_{x_k,x_{k+1}}=1, \ k\geq
0\}
$$
which is the corresponding subshift space, and let $\sigma$ be the
shift transformation on $\Sigma_A$. We equip $\Sigma_A$ with a
metric $d_f$ depending on the dynamics which is defined as follows.
First for $i,j\in I,\ i\neq j$ let $\kappa(i,j)$ be the integer such
that $|a_i-a_j|_p=p^{-\kappa(i,j)}$.  It is clear that
$\kappa(i,j)<-\log(r)$, where $r$ is the radius of the balls at the
beginning of this section. By the ultrametric inequality, we have
$$
|x-y|_p=|a_i-a_j|_p\ \ \ i\neq j,\ \forall x\in B_r(a_i), \forall
y\in B_r(a_j)
$$
For $x=(x_0,x_1,\dots,x_n,\dots)\in\Sigma_A$ and
$y=(y_0,y_1,\dots,y_n,\dots)\in\Sigma$, define
$$
d_f(x,y)=\left\{\begin{array}{ll}
p^{-\tau_{x_0}-\tau_{x_1}-\cdots-\tau_{x_{n-1}}-\kappa(x_{n},y_{n})}&, \mbox{ if }n\neq0\\
p^{-\kappa(x_0,y_0)}&, \mbox{ if }n=0
\end{array}\right.
$$
where $n=n(x,y)=\min\{i\geq0: x_i\neq y_i\}$. It is clear that $d_f$
defines the same topology as the classical metric which is defined
by $d(x,y)=p^{-n(x,y)}$.

\begin{theorem}\label{xit}\cite{FL2} Let $(X,J_f,f)$ be a transitive $p$-adic weak repeller with incidence matrix $A$.
Then the dynamics $(J_f,f,|\cdot|_p)$ is isometrically conjugate to
the shift dynamics $(\Sigma_A,\sigma,d_f)$.
\end{theorem}

\subsection{$p$-adic measure}
Let $(X,\mathcal B)$ be a measurable space, where $\mathcal B$ is an algebra of
subsets of $X$. A function $\mu:\mathcal B\to \mathbb Q_p$ is said to be a {\it
$p$-adic measure} if for any $A_1,\dots,A_n\in\mathcal B$ such that
$A_i\cap A_j=\varnothing$ ($i\neq j$) the following equality holds
$$
\mu\bigg(\bigcup_{j=1}^{n} A_j\bigg)=\sum_{j=1}^{n}\mu(A_j).
$$

A $p$-adic measure is called a {\it probability measure} if
$\mu(X)=1$.  One of the important condition (which was already
invented in the first Monna--Springer theory of non-Archimedean
integration \cite{Mona}) is boundedness, namely a $p$-adic
probability measure $\mu$ is called {\it bounded} if
$\sup\{|\mu(A)|_p : A\in \mathcal B\}<\infty $. We pay attention to an
important special case in which boundedness condition by itself
provides a fruitful integration theory (see for example
\cite{Kh07}). Note that, in general, a $p$-adic probability
measure need not be bounded \cite{Ro}. For more detail information
about $p$-adic measures we refer to \cite{K3,KhN,Ro}.

\subsection{Cayley tree}
Let $\Gamma^k_+ = (V,L)$ be a semi-infinite Cayley tree of order
$k\geq 1$ with the root $x^0$ (whose each vertex has exactly $k+1$
edges, except for the root $x^0$, which has $k$ edges). Here $V$
is the set of vertices and $L$ is the set of edges. The vertices
$x$ and $y$ are called {\it nearest neighbors} and they are
denoted by $l=\langle x,y\rangle $ if there exists an edge connecting them. A
collection of the pairs $\langle x,x_1\rangle ,\dots,\langle x_{d-1},y\rangle $ is called a
{\it path} from the point $x$ to the point $y$. The distance
$d(x,y), x,y\in V$, on the Cayley tree, is the length of the
shortest path from $x$ to $y$.

Denote
$$
W_{n}=\left\{ x\in V: d(x,x^{0})=n\right\}, \ \
V_n=\bigcup_{m=0}^nW_{m}, \ \ L_{n}=\left\{\langle x,y\rangle \in L: x,y\in V_{n}\right\}.
$$
The set of direct successors of $x\in W_n$ is defined by
$$
S(x)=\left\{ y\in W_{n+1}: d(x,y)=1\right\}.
$$
Observe that any vertex $x\neq x^{0}$ has $k$ direct successors
and $x^{0}$ has $k+1$.

Now we are going to introduce a coordinate structure in $\Gamma_+^k$.
Every vertex $x$ (except for $x^{0}$) of $\Gamma_+^k$ has coordinates
$(i_1,\dots,i_n)$, here $i_m\in\{1,\dots,k\},\ 1\leq m\leq n$ and
for the vertex $x^0$ we put $(0)$ (see Figure 1). Namely, the
symbol $(0)$ constitutes level $0$ and the sites $i_1,\dots,i_n$
form level $n$ of the lattice. In this notation for $x\in\Gamma_+^k,\
x=(i_1,\dots,i_n)$ we have
$$
S(x)=\{(x,i): 1\leq i\leq k\},
$$
here $(x,i)$ means that $(i_1,\dots,i_n,i)$.

\begin{figure}
\begin{center}
\includegraphics[width=10.07cm]{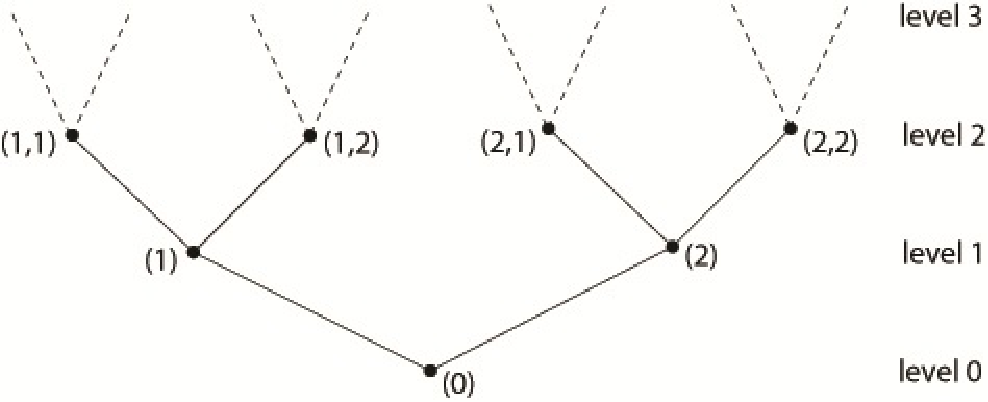}
\end{center}
\caption{The first levels of $\Gamma_+^2$} \label{fig1}
\end{figure}

Let us define on $\Gamma_+^k$ a binary operation $\circ:\Gamma^k_+\times\Gamma_+^k\to\Gamma_+^k$
as follows. For any two elements $x=(i_1,\dots,i_n)$ and $y=(j_1,\dots,j_m)$ put
$$
x\circ y=(i_1,\dots,i_n)\circ(j_1,\dots,j_m)=(i_1,\dots,i_n,j_1,\dots,j_m)
$$
and
$$
y\circ x=(j_1,\dots,j_m)\circ(i_1,\dots,i_n)=(j_1,\dots,j_m,i_1,\dots,i_n).
$$
By means of the defined operation, $\Gamma_+^k$ becomes a
noncommutative semigroup with a unit. Using this semigroup
structure one defines translations $\tau_g:\Gamma_+^k\to\Gamma_+^k,\
g\in\Gamma_k$ by
$$
\tau_g(x)=g\circ x.
$$
Similarly, by means of $\tau_g$, one can define translation $\tilde\tau_g:L\to L$ of $L$. Namely,
$$
\tilde\tau_g(\langle{x,y}\rangle)=\langle{\tau_g(x),\tau_g(y)}\rangle.
$$

Let $G\subset\Gamma_+^k$ be a sub-semigroup of $\Gamma_+^k$ and $h:L\to\mathbb Q_p$ be a function defined on $L$.
We say that $h$ is a $G$-{\it periodic} if $h(\tilde\tau_g(l))=h(l)$ for all $g\in G$ and $l\in L$.
A $\Gamma^k_+$-periodic function is called {\it translation-invariant}. Put
$$
G_m=\left\{x\in\Gamma_+^k: d(x,x^0)\equiv0(\operatorname{mod }m)\right\},\ \ \ m\geq2.
$$

One can check that $G_m$ is a sub-semigroup with a unit.

\section{$p$-adic $\lambda$ model and its $p$-adic quasi Gibbs measures}
\label{sec:3}
In this section, we consider the $p$-adic $\lambda$-model where the spin
takes values in the set $\Phi=\{-1,+1\}$, ($\Phi$ is called a {\it
state space}) and is assigned to the vertices of the tree
$\Gamma^k_+=(V,L)$. The interaction function $\lambda$ is specified later on. A configuration $\sigma$ on $V$ is then defined as a
function $x\in V\to\sigma(x)\in\Phi$; in a similar manner one defines
configurations $\sigma_n$ and $\omega$ on $V_n$ and $W_n$, respectively.
The set of all configurations on $V$ (resp. $V_n$, $W_n$)
coincides with $\Omega=\Phi^{V}$ (resp. $\Omega_{V_n}=\Phi^{V_n},\
\ \Omega_{W_n}=\Phi^{W_n}$). One can see that
$\Omega_{V_n}=\Omega_{V_{n-1}}\times\Omega_{W_n}$. Using this, for given
configurations $\sigma_{n-1}\in\Omega_{V_{n-1}}$ and $\omega\in\Omega_{W_{n}}$
we define their concatenations  by
$$
(\sigma_{n-1}\vee\omega)(x)= \left\{
\begin{array}{ll}
\sigma_{n-1}(x), \ \ \textrm{if} \ \  x\in V_{n-1},\\
\omega(x), \ \ \ \ \ \ \textrm{if} \ \ x\in W_n.\\
\end{array}
\right.
$$
It is clear that $\sigma_{n-1}\vee\omega\in \Omega_{V_n}$.

Assume a function $\lambda:\Phi\times\Phi\to
{\mathbb Z}$ is given. Then the Hamiltonian $H_n:\Omega_{V_n}\to\mathbb Z$ of the
$p$-adic $\lambda$-model is defined by
\begin{equation}\label{ham1}
H_{n}(\sigma)=\sum\limits_{\langle  x,y \rangle \in L_{n}}\lambda(\sigma(x),\sigma(y))
\end{equation}

\begin{remark} This model first has been considered in \cite{KM}. In
the real setting such kind of model was studied in \cite{R1}. We
remark that if one takes $\lambda(u,v)=Nuv$ for some integer $N$, then
the model \eqref{ham1} reduces to the well-known Ising model (see
\cite{GMR,KM,Mq}).
\end{remark}

Let $\rho\in\mathbb Q_p\setminus\{-1,0,1\}$ and assume that ${\bf h}:x\in
{V\setminus\{x^{(0)}\}}\to \mathbf{h}_x\in\mathbb Q_p^\Phi$ be a mapping, i.e.
${\bf h}_x=(h_{x,-1},h_{x,1}),$ where $h_{x,\pm 1}\in\mathbb{Q}_p$. Given
$n\in\mathbb N$, let us consider a $p$-adic probability measure
$\mu^{(n)}_{{\bf h},\rho}$ on $\Omega_{V_n}$ defined by

\begin{equation}\label{measure1}
\mu_{{\bf h},\rho}^{(n)}(\sigma)=\frac{1}{Z_{n,\rho}^{({\bf h})}}\rho^{H_{n}(\sigma)}
{\underset{x\in W_{n}}{\prod}h_{x,\sigma(x)}}
\end{equation}
Here, $\sigma\in\Omega_{V_n}$, and $Z_{n,\rho}^{({\bf h})}$ is the corresponding
normalizing factor called a {\it partition function} given by

\begin{equation}\label{partition}
Z_{n,\rho}^{({\bf h})}=\sum_{\sigma\in\Omega_{V_n}}\rho^{H_{n}(\sigma)}{\underset{x\in
W_{n}}{\prod}h_{x,\sigma(x)}}.
\end{equation}

\begin{remark}\label{remrem}
Note that in general, $Z_{n,\rho}^{({\bf h})}$ could be zero for some ${\bf h}$. In this case, formally, we may
assume that $\mu_{{\bf h}, \rho}^{(n)}(\sigma)=\infty$ for all $\sigma\in\Omega_{V_n}$. However, such kind of measures are not interested.
Hence, when it occurs we say that for that ${\bf h}$ there is no measure.
\end{remark}

We recall \cite{M12} that one of the central results of the theory
of probability concerns a construction of an infinite volume
distribution with given finite-dimensional distributions, which is
a well-known {\it Kolmogorov's extension Theorem} \cite{Sh}.
Recall that a $p$-adic probability measure $\mu_{\bf h}$ on $\Omega$ is
\textit{compatible} with defined ones $\mu_{{\bf h},\rho}^{(n)}$ if one has
\begin{equation}\label{CM}
\mu_{\bf h}(\sigma\in\Omega: \sigma|_{V_n}=\sigma_n)=\mu^{(n)}_{{\bf h},\rho}(\sigma_n), \ \ \
\textrm{for all} \ \ \sigma_n\in\Omega_{V_n}, \ n\in\mathbb{N}.
\end{equation}

The existence of the measure $\mu$ is guaranteed by the $p$-adic
Kolmogorov's Theorem \cite{GMR,KL}. Namely, if the measures
$\mu_{{\bf h},\rho}^{(n)}$, $n\geq 1$ satisfy the {\it compatibility
condition}, i.e.
\begin{equation}\label{comp}
\sum_{\omega\in\Omega_{W_n}}\mu^{(n)}_{{\bf h},\rho}(\sigma_{n-1}\vee\omega)=\mu^{(n-1)}_{{\bf h},\rho}(\sigma_{n-1}),
\end{equation}
for any $\sigma_{n-1}\in\Omega_{V_{n-1}}$, then there is a unique measure
$\mu$ on $\Omega$ with (\ref{CM})\footnote{In the real case, when the
state space is compact, then the existence follows from the
compactness of the set of all probability measures (i.e.
Prohorov's Theorem). When the state space is non-compact, then
there is a Dobrushin's Theorem \cite{Dob1,Dob2} which gives a
sufficient condition for the existence of the Gibbs measure for a
large class of Hamiltonians.}.

Now following \cite{M12} if for some function ${\bf h}$ the measures
$\mu_{{\bf h},\rho}^{(n)}$ satisfy the compatibility condition, then there
is a unique $p$-adic probability measure, which we denote by
$\mu_{{\bf h},\rho}$, since it depends on ${\bf h}$ and $\rho$. Such a measure
$\mu_{\bf h}$ is said to be {\it a generalized $p$-adic quasi Gibbs
measure} corresponding to the $p$-adic $\lambda$-model. By $Q\mathcal G(H)$ we
denote the set of all generalized $p$-adic quasi Gibbs measures
associated with the Hamiltonian $H$ \eqref{ham1}. If there are at
least two distinct generalized $p$-adic quasi Gibbs measures
$\mu,\nu\in Q\mathcal G(H)$ such that $\mu$ is bounded and $\nu$ is
unbounded, then we say that {\it a phase transition} occurs. By
another words, one can find two different functions ${\bf s}$ and ${\bf h}$
defined such that there exist the corresponding measures
$\mu_{{\bf s},\rho}$ and $\mu_{{\bf h},\rho}$, for which one is bounded, another
one is unbounded. If there are two different
functions ${\bf s}$ and ${\bf h}$ defined on $V\setminus\{x^{(0)}\}$ such that there exist
the corresponding measures $\mu_{{\bf s},\rho}$, $\mu_{{\bf h},\rho}$, and
they are bounded, then we say there is a \textit{quasi phase
transition}.

Note that some comparison of these phase transitions with real
counterparts was highlighted in \cite{M12}.

\begin{remark} Note that if one takes $\rho=p$,
then the generalized $p$-adic quasi Gibbs measure reduces to the
$p$-adic quasi Gibbs measure (see \cite{Mq}). If one has
$\rho\in\mathcal E_p$ and $\bf h_x\in \mathcal E_p$ (for all $x\in V)$, then the defined measure
reduces to the $p$-adic Gibbs measure (see \cite{KM,KMR}).
\end{remark}

One can prove the following theorem.

\begin{theorem}\label{compatibility}
The measures $\mu^{(n)}_{{\bf h},\rho}$, $ n=1,2,\dots$ (see
(\ref{measure1})), associated with $\lambda$-model (\ref{ham1}),
satisfy the compatibility condition (\ref{comp}) if and only if
for any $x\in V$ the following equation holds:
\begin{equation}\label{canonic3}
\tilde h_x=\prod_{y\in S(x)}\left(\frac{\rho^{\lambda(1,1)}\tilde h_y+\rho^{\lambda(1,-1)}}
{\rho^{\lambda(-1,1)}\tilde h_y+\rho^{\lambda(-1,-1)}}\right),
\end{equation}
where $\tilde h_x=\frac{h_{1,x}}{h_{-1,x}}$.
\end{theorem}

The proof can be proceeded by the same argument as in \cite{KM}.

According to Theorem \ref{compatibility}, the problem of describing
the generalized $p$-adic quasi Gibbs measures is reduced to the
description of solutions of the functional equations
(\ref{canonic3}). Note that in \cite{M07,MA15} some general methods to
solve the above kind of equations have been proposed.

Recall that a function ${\bf h}$ is \textit{translation-invariant} if ${\bf h}_x={\bf h}_y$
for all $x,y\in V$. Then the corresponding $p$-adic measure is also called
\textit{translation-invariant}. If ${\bf h}$ is translation-invariant then
(\ref{canonic3}) reduces to the equation $F(h)=h$, where $h=\tilde h_x$ for all $x$. Here
\begin{equation}\label{func}
F(h)=\left(\frac{\rho^{\lambda(1,1)}h+\rho^{\lambda(1,-1)}}
{\rho^{\lambda(-1,1)}h+\rho^{\lambda(-1,-1)}}\right)^k.
\end{equation}

\section{Dynamical system for the $p$-adic Ising Model}

In this section, we are going to consider a very particular case of the
$\lambda$-model corresponding to the Ising case.  In this situation,
we have $\lambda(u,v)=Nuv$. Then the associated dynamical system reduces to the following one:
\begin{equation}\label{func}
f_\theta(x)=\left(\frac{\theta x+1}{x+\theta}\right)^k,\ \ \ \theta=\rho^{2N}\in\mathbb{Q}_p.
\end{equation}

We recall that the function (\ref{func}) is called the
\textit{Ising-Potts mapping}.

One can observe that to find $H_m$-periodic $p$-adic quasi Gibbs
measures for the model it is enough to solve the following system:
\begin{equation}\label{perh}
\begin{array}{ll}
h_i=f_\theta(h_{i+1}),\ \ i=\overline{1,m-1}\\
h_m=f_\theta(h_1).
\end{array}
\end{equation}
The last one is equivalent to find $m$-periodic points of the
function $f_\theta$. Hence, the existence of periodic orbits of the
function (\ref{func}) yields the existence of $H_m$-periodic
$p$-adic quasi Gibbs measures.

In what follows, by $Fix(f_{\theta})$ we denote the set of all fixed points of (\ref{func}).
The following result shows a relation between the set $Fix(f_{\theta})$  and
the set of all translation-invariant generalized $p$-adic Gibbs measures for Ising model.

\begin{proposition}\label{pro_asosiymain}
Let $h\neq-1$ be a fixed point of (\ref{func}). Then $\mu_{h}:=\mu_{{\bf h},\rho}$
is a translation-invariant generalized
$p$-adic Gibbs measure for
Ising model, where ${\bf h}_x=(1,h)$ for every $x\in V$. Moreover,
\begin{equation}\label{mu_h=z_h}
\mu_{h}(\left\{\sigma\in\Omega: \sigma|_{V_n}\equiv\sigma_n\right\})=\frac{\rho^{H_n(\sigma)}h^{\sum_{x\in W_n}\delta_{1\sigma(x)}}}{\left(\rho^{-N}h+\rho^N\right)^{\frac{k(k^n-1)}{k-1}}(h+1)},\ \ \ \ \ \forall n\in\mathbb N,
\end{equation}
where $\delta_{ij}$ is the Kronecker's symbol.
\end{proposition}

In order to study the existence phase transition
(or strong, quasi types of phase transition) for Ising model we need to know the boundedness
(or unboundedness) of the given measures. Next result is crucial in our further investigations.

For the function $f_{\theta}$ the following statements hold true:
\begin{itemize}
\item[(A)] \ \  if $|\theta|_p\leq1$ then $Fix(f_{\theta})\subset\mathbb Z_p^*$;
\item[(B)] \ \ if $|\theta|_p>1$ then $Fix(f_{\theta})\subset\bigcup_{t\in\{-1,0,1\}}\theta^{kt}\mathbb Z_p^*$.
\end{itemize}

These facts allow us to prove the following result.

\begin{theorem}\label{thmbound} Let $\rho^{2N}\in\mathbb Q_p\setminus\{-1,0,1\}$ and $\mu_h$ be a translation-invariant generalized
$p$-adic Gibbs measure for the Ising model. Then, the following are true:
\begin{itemize}
\item[(I)] \ \  Let $|\rho|_p\neq1$. Then $\mu_h$ is bounded;
\item[(II)] \ \ Let $|\rho|_p=1$. Then $\mu_h$ is unbounded if and only if $0<|h+\rho^{2N}|_p<1$.
\end{itemize}
\end{theorem}

\begin{corollary}\label{corp=2}
Let $p=2$ and $\mu_h$ be a translation-invariant generalized
$p$-adic Gibbs measure for the Ising model. Then $\mu_h$ is unbounded if and only if $|\rho|_2=1$.
\end{corollary}

From Theorem \ref{thmbound} and Corollary \ref{corp=2}, we immediately get the following result.

\begin{corollary}\label{corunbound}
On the set of all translation-invariant generalized $p$-adic Gibbs measures for the Ising model, a
phase transition does not occur if $p=2$ or $|\rho|_p\neq1$ for $p>2$.
\end{corollary}

We recall that an integer $a\in\mathbb Z$ is called the
\textit{$k$th residue modulo} $p$ if the congruent equation
$x^k\equiv a\ (\operatorname{mod }p)$ has a solution $x\in\mathbb Z$.

Denote $\mathbb F_p=\mathbb Z/p\mathbb Z$. Recall that $\sqrt[k]{-1}$ exists in $\mathbb F_p$
whenever $-1$ is the $k$th residue of module $p$. Otherwise, it is
said that $\sqrt[k]{-1}$ does not exist in $\mathbb F_p$. Let
$\mathcal N_{k,p}$ be the number of solutions of
$x^k\equiv-1(\operatorname{mod }p)$ in $\mathbb F_p$. It is known
\cite{Rosen} that $\sqrt[k]{-1}$ exists in $\mathbb F_p$ if and only
if $\frac{p-1}{(k,p-1)}$ is even. Moreover, in this case, one has
that $\mathcal N_{k,p}=(k, p-1)$, here $(k,p-1)$ means the common
divisor of the numbers $k$ and $p-1$. Similarly, we say that
$\sqrt[k]{-1}$ exists in $\mathbb Q_p$ whenever the equation
$x^k=-1$ is solvable in $\mathbb Q_p$. Otherwise, it is said that
$\sqrt[k]{-1}$ does not exist in $\mathbb Q_p$. It was shown
\cite{MK20,MS13} that $\sqrt[k]{-1}$ exists in $\mathbb Q_p$ if and only
if $\sqrt[q]{-1}$ exists in $\mathbb F_p$, where $k=qp^s$ and $(q,
p)=1$ with $s\geq0$.
We point out that the description of this set has been carried out in \cite{MK20,MS13,MOS}.

By $TIG\mathcal G_pM$ we denote the set of all translation-invariant generalized $p$-adic Gibbs measures for the Ising model.

\begin{theorem}\label{TII}\cite{MK21} Let $k\geq2$.  Then for the Ising model on $\Gamma_+^k$ the following statements hold:
\begin{itemize}
\item[(i)] \ If $|\rho^N|_p>1$, then $card\left(TIG\mathcal G_pM\right)\geq3$. Moreover, if
$|\rho^{-N}|_p<|k-1|_p$ then
$$
card\left(TIG\mathcal G_pM\right)=\left\{
\begin{array}{ll}
\mathcal N_{p,k}+2, & \mbox{if k is even};\\
\mathcal N_{p,k}+1, & \mbox{if k is odd}.
\end{array}
\right.
$$

\item[(ii)]\ \ If $|\rho^N|_p<1$, then $card\left(TIG\mathcal G_pM\right)\geq1$. Moreover, if  $|\rho^{N}|_p<|k+1|_p$ then
$$
card\left(TIG\mathcal G_pM\right)=\left\{
\begin{array}{ll}
\mathcal N_{p,k}+1, & \mbox{if k is even};\\
\mathcal N_{p,k}, & \mbox{if k is odd}.
\end{array}
\right.
$$
\end{itemize}
\end{theorem}

We notice that certain properties of the generalized $p$-adic
Gibbs measures of the Ising model have been carried out in \cite{MRTM23,RKT19,RT19,RT23}.

We set
\begin{equation}\label{solx2n}
\mbox{Sol}_p(x^k+1):=\left\{\xi\in\mathbb F_p\setminus\{p-1\}:
\xi^k+1\equiv0(\operatorname{mod }p)\right\},
\end{equation}
and $\kappa_p:=|\mbox{Sol}_p(x^k+1)|$, where $|A|$ stands for the
cardinality of a set $A$.

Note that $\mathcal N_{k,p}=\kappa_p$ if $k$ is even and $\mathcal
N_{k,p}=\kappa_p+1$ if $k$ is odd.

\begin{proposition}\label{propfix} Let $p\geq3$ and $k\geq1$. Assume that $\rho\in\mathcal{E}_p$.
Then $x_0=1$ is an attracting fixed point of $f_\theta$ and
$$
\mathcal A(x_0)=\bigcup_{n\geq0}f_\theta^{-n}(\mathcal E_p).
$$
Moreover, $\mathcal A(x_0)=\mbox{Dom}(f_\theta)$ if $\mbox{Sol}_p(x^k+1)=\varnothing$.
\end{proposition}

Now we assume that $\mbox{Sol}_p(x^k+1)\neq\varnothing$ and denote
\begin{equation}\label{x_i}
x_i=-1+(\theta-1)\eta_i,\ \ \ i=\overline{1,\kappa_p}
\end{equation}
where $\eta_i\in\mathbb \{0,1,\dots,p-1\}$ is a solution of
$\eta_i(\xi_i-1)+\xi_i+1\equiv0(\operatorname{mod }p)$ for a given
$\xi_i\in\mbox{Sol}_p(x^k+1)$.

\begin{remark} \begin{itemize}
\item[(i)] We note that from the definition of the set
$\mbox{Sol}_p(x^k+1)$, we infer that $\xi\neq p-1$ for any
$\xi\in\mbox{Sol}_p(x^k+1)$, which implies that $\eta_i\neq0$.
Moreover, due to $p\geq3$ we have $\eta_i\neq p-1$.

\item[(ii)] We point out that the solvability of the linear congruence
$(\xi-1)x+\xi+1\equiv0(\operatorname{mod }p)$ for any $\xi\in\mathbb
F_p$ follows from the solvability of the Diophantine Equation
$py-(\xi-1)x=\xi+1$ for any prime $p\geq3$.
\end{itemize}
\end{remark}

In \cite{MK17} we have established the following result.

\begin{theorem}\label{mainres} Let $p\geq3,\ \kappa_p\geq2$, and  $\rho\in\mathcal{E}_p$ with $|\theta-1|_p<|k|_p$.
Assume that $$ X=\bigcup_{i=1}^{\kappa_p} B_r(x_i), \ \ \
r=|p(\theta-1)|_p,$$ where $x_i$ is defined by (\ref{x_i}). Let
$f_\theta: X\to\mathbb Q_p$ be a function defined by (\ref{func})
then the dynamics $(J_{f_\theta},f_\theta,|\cdot|_p)$ is
isometrically conjugate to the full shift
$(\Sigma_{\kappa_p},\sigma,d_f)$.
\end{theorem}

This theorem allows us to establish the next important result about the existence of periodic points of $f_\theta$.

\begin{theorem}\label{mainres0}
Let $p\geq3$ and $\frac{p-1}{(k,p-1)}$ be even. If $(k,p-1)\geq2$,
and $\rho\in\mathcal E_p$ with $|\theta-1|_p<|k|_p$, then there exist infinitely many
$H_m$-periodic $p$-adic quasi Gibbs measures for the $p$-adic Ising
model on a Cayley tree of order $k$.
\end{theorem}

Chaotic behavior in various types of dynamical systems has been studied
in \cite{MAD17, MFKh162, MK161, X}. Explicit constructions of periodic
Gibbs measures have been developed in \cite{Khak1,Khak11,Khak2,RKT19, RT22, RK2, RST22}.

Now, considering Theorem \ref{TII}, we can establish similar result
like Theorem \ref{mainres0} when $\rho\notin\mathcal E_p$.  Such a kind of
finding should be based on the next problem.

\begin{problem}
Assume $\rho\notin\mathcal E_p$, and consider the mappings $f_\theta$. Investigate its chaotic behavior.
\end{problem}

In \cite{RA241,RA24,RA25,RTS24,T24} weakly periodic generalized $p$-adic Gibbs measures are discussed for the Ising model.

According to Theorem \ref{mainres0}, there are infinitely many periodic
$p$-adic quasi Gibbs measures. Therefore, we can formulate the next problem

\begin{problem} Assume $\rho\in\mathcal E_p$. Are there infinitely many weakly
periodic $p$-adic quasi Gibbs measures which are not periodic?
Investigate the same question for the case  $\rho\notin\mathcal E_p$ as well.
\end{problem}

\section{Dynamical system and the existence of generalized $p$-Adic quasi Gibbs Measures}

In this section we consider the $\lambda$-model (\ref{ham1}) over the
Cayley tree of order two, i.e. $k=2$. Main aim of this section is
to establish the existence of generalized $p$-adic quasi Gibbs
measures by analyzing the equation (\ref{canonic3}). In the
sequel, we will consider a case when $|\rho|_p<1$ and $p\geq 3$.
Note that the case $\rho\in\mathcal E_p$ has been studied in
\cite{KM,KMR,MD15}.

To solve the equation (\ref{canonic3}), in general, is very
complicated. Therefore,
 let us first restrict ourselves to
the description of translation-invariant solutions of
(\ref{canonic3}). More exactly, we suppose that $h_{x}:=h$ for all
$x\in V$.

\subsection{Regime $|\rho|_p<1$} Then from \eqref{canonic3}, we find
\begin{equation}\label{nnn}
h=\left(\frac{Ah+B} {Ch+D}\right)^2,
\end{equation}
where $A=\rho^{\lambda(1,1)}$, $B=\rho^{\lambda(1,-1)}$,
$C=\rho^{\lambda(-1,1)}$,
$D=\rho^{\lambda(-1,-1)}$.

The last equation can be reduced to the following one:
\begin{equation}\label{nnn31}
h=\frac{Ah^2+B} {Ch^2+D}.
\end{equation}

Let us assume that $|A|_p,|C|_p<1$, $B=D=1$, i.e.
\[
\lambda(1,1),\lambda(-1,1)\in\mathbb N,\ \lambda(1,-1)=\lambda(-1,-1)=0.
\]
In what follows,
we will assume that $|A|_p\neq|C|_p$, otherwise one finds $A=C$
and correspondingly equation (\ref{nnn31}) becomes trivial.

Let us denote
$$
S=\{x\in\mathbb Q_p:\ \ |x|_p=1\}.
$$

\begin{lemma}\label{parti-g1}
Let $p\geq3$. Suppose $|A|_p,|C|_p<1$. Let $f$ be given by
\begin{equation}\label{g11}
f(x)=\frac{Ax^2+1}{Cx^2+1}.
\end{equation}
Then $f(S)\subset S$ and
$$|f(x)-f(y)|_{p}\leq|A-C||x-y|_{p},$$
 for all $x,y\in S$.
\end{lemma}

Now we can formulate the following proposition on the fixed points
of $f$.

\begin{theorem}\cite{MD15}\label{g1-fix} Let $|A|_p,|C|_p<1$ with $|A|_p\neq |C|_p$, and $f$
be given by (\ref{g11}). Then the following statements hold.
\begin{enumerate}
\item[(i)] The function $f$ has a unique fixed point $x_{0}$ in
$\mathcal E_{p}$, which is attractive. \item[(ii)] Assume that $|A|_p^2<|C|_p$. Then the
function $f$ has at most two fixed points $x_{1},x_{2}$ different
from $x_0$ if and only if $\sqrt{-C}$ exists; Moreover, $x_{1,2}$ are repelling, and
\begin{equation}\label{x12-1}
|x_{1,2}|_p=\frac{1}{\sqrt{|C|_p}}.
\end{equation}

\item[(iii)] Assume that $|A|_p^2>|C|_p$. Then the function $f$
has two fixed points $x_{1},x_{2}$ different from $x_0$. Moreover, $x_{1}$ is attractive and $x_2$ is neutral, and
\begin{equation}\label{x12-2}
|x_{1}|_p=\frac{|A|_p}{|C|_p}, \ |x_{2}|_p=\frac{1}{|A|_p}.
\end{equation}
\end{enumerate}
\end{theorem}

According to Theorem \ref{compatibility} the solutions $x_{0}$,
$x_{1}$ and $x_{2}$ (if they exist) generate generalized $p$-adic
quasi Gibbs measures $\mu_{0}$, $\mu_{1}$ and $\mu_{2}$,
respectively. Hence, we can formulate the following result.

\begin{theorem}\label{exist-11}
Let $p\geq3$, $|\rho|_p<1$. Assume that for the function $\lambda$ one
has
\begin{equation}\label{lll}
 \lambda(1,1),\ \lambda(-1,1)>0,  \ \ \lambda(1,-1)=\lambda(-1,-1)=0.
\end{equation}
 Then for the $\lambda$-model (\ref{ham1}) on the Cayley tree of order two, the
following assertions hold:
\begin{enumerate}
\item[(i)] There exists a transition-invariant generalized
$p$-adic quasi Gibbs Measure $\mu_{0}$.

\item[(ii)] If
$$2\lambda(1,1)>\lambda(-1,1),
$$ then there are three
transition-invariant generalized $p$-adic quasi Gibbs measures
$\mu_{0}$, $\mu_{1}$ and $\mu_{2}$ if and only if
$\sqrt{-\rho^{\lambda(-1,1)}}$ exists.

\item[(ii)] If
$$2\lambda(1,1)<\lambda(-1,1),$$ then there are three
transition-invariant generalized $p$-adic quasi Gibbs measures
$\mu_{0}$, $\mu_{1}$ and $\mu_{2}$.
\end{enumerate}
\end{theorem}

\begin{remark} We note that dynamical behavior of certain fractional $p$-adic
dynamical systems have been studied in \cite{ARS,FL3,KM2,MR3,RL}.
Therefore, further studies of dynamical behavior of the dynamical
system (\ref{g11}) will be carried out elsewhere.
\end{remark}

We stress that the measure $\mu_0$ indeed is unique $p$-adic Gibbs
measure for the model.

\begin{theorem}\label{SPT}
Let $p\geq3$, $|\rho|_p<1$. Assume that for the function $\lambda$ one
has
\begin{equation*}\label{lll}
 2\lambda(1,1)>\lambda(-1,1),  \ \ \lambda(1,-1)=\lambda(-1,-1)=0.
\end{equation*}
and $\sqrt{-\rho^{\lambda(-1,1)}}$ exists.
 Then, there exist a phase transition for the
$\lambda$-model (\ref{ham1}) on the Cayley tree of order two.
\end{theorem}

\begin{remark} This result confirms that if the dynamical system associated with a model
 has at least
two repelling fixed points, then the model exhibits a phase
transition. We stress that the considered $\lambda$-model has a
stronger phase transition (see \cite{MD15}).
\end{remark}

\begin{remark} If one takes $\rho=p$ and $\lambda(-1,1)=2m$ for some
$m\in\mathbb N$, then $\sqrt{-p^{2m}}$ exists if and only if $p\equiv
1(\textrm{mod}\ 4)$.
\end{remark}

\begin{theorem}\label{QPT}
Let $p\geq3$, $|\rho|_p<1$. Assume that for the function $\lambda$ one
has
\begin{equation}\label{lll-1}
 2\lambda(1,1)<\lambda(-1,1),  \ \ \lambda(1,-1)=\lambda(-1,-1)=0.
\end{equation}
 Then, there exist a quasi phase transition for the
$\lambda$-model (\ref{ham1}) on the Cayley tree of order two.
\end{theorem}

\subsection{Regime $\rho\in\mathcal E_p$}

In this regime, we rewrite
(\ref{nnn}) as follows
\begin{equation}\label{nnn2}
h=\left(a\bigg(\frac{bh+1}{h+c}\bigg)\right)^2.
\end{equation}
where $a=\rho^{\lambda(1,-1)-\lambda(-1,1)}$, $b=\rho^{\lambda(1,1)-\lambda(1,-1)}$,
$c=\rho^{\lambda(-1,-1)-\lambda(-1,1)}$. In the considered regime we have
$a,b,c\in\mathcal E_p$.

One can establish that the last equation reduces to
\begin{equation}\label{nnn31}
h=a\bigg(\frac{bh^2+1}{h^2+c}\bigg).
\end{equation}

\begin{remark} In the sequel, for the sake of simplicity, we will assume that
$|b-1|_p+|c-1|_p\neq 0$, otherwise, (\ref{nnn31}) becomes trivial,
and it defines a unique generalized $p$-adic quasi Gibbs measure.
Therefore, in what follows, we will always assume that $b,c\neq 1$.
\end{remark}

Let us denote
\begin{equation}\label{gg}
g(u)=a\left(\frac{bu^{2}+1}{u^{2}+c}\right).
\end{equation}

Now, we can obtain the following proposition about fixed points of
$g$.

\begin{proposition}\label{g-fix} Let $a,b,c\in\mathcal E_{p}$ with
$|b-1|_p+|c-1|_p\neq 0$, and $g$ be given by (\ref{gg}). Then the following statements hold:
\begin{enumerate}
\item[(i)] The function $g$ has a unique fixed point $x_{0}$ in
$\mathcal E_{p}$.
\item[(ii)] The function $g$ has at most two fixed points
$x_{1},x_{2}$ different from $x_0$
if and only if $\sqrt{-1}$ exists.

\item[(iii)] Let $x_{1},x_{2}$ be two fixed points of $g$ different from
$x_0$. Then $x_{1},x_{2}\in\mathbb{Q}_p\setminus\mathcal{E}_p$.
\end{enumerate}
\end{proposition}

 Now we can
 obtain the following result about the existence of generalized $p$-adic quasi Gibbs measures.

\begin{theorem}\label{exist}
Let $p\geq3$, $\rho\in\mathcal E_p$ and $|b-1|_p+|c-1|_p\neq 0$. Then for the
$\lambda$-model (\ref{ham1}) on the Cayley tree of order two the
following assertions hold:
\begin{enumerate}
\item[(i)] There exists a transition-invariant generalized $p$-adic quasi
Gibbs measure $\mu_{0}$.

\item[(ii)] There are three
transition-invariant generalized $p$-adic quasi Gibbs measures
$\mu_{0}$, $\mu_{1}$ and $\mu_{2}$ if and only if $p\equiv
1(\textrm{mod}\ 4)$.
\end{enumerate}
\end{theorem}

\begin{theorem}\label{PT1}
Let $p\geq3$,  $\rho\in\mathcal E_p$ and $|b-1|_p+|c-1|_p\neq 0$. Assume that
$p\equiv 1(\textrm{mod}\ 4)$. Then, there exists a phase transition
for the $p$-adic $\lambda$-model (\ref{ham1}) on the Cayley tree of order
two.
\end{theorem}

\begin{remark} Note that in \cite{KM,KMR,Khak1} a similar theorem has been
proved for the $p$-adic Ising model on the Cayley tree.
\end{remark}

Now, keeping in mind Theorem \ref{mainres}, we can formulate the following problems.

\begin{problem} Let us consider the mappings $f$ and $g$ given
by (\ref{g11}) and (\ref{gg}), respectively. Under what conditions of
$A,C$ and $a,b,c\in\mathcal E_p$, respectively, the functions $f$ and $g$ have chaotic behavior.
\end{problem}

\begin{problem} Investigate periodic and weakly periodic
$p$-adic quasi Gibbs measures for the $\lambda$-model on
the Cayley tree of order $k\geq 2$.
\end{problem}

\section*{Acknowledgments}
The first named author (F.M.) thanks to the UAEU UPAR Grant No. G00004962 for support.
The authors are also greatly indebted to anonymous referees whose constructive comments/suggestions
substantially contributed to improving the quality and presentation of this paper

%
%
%

\end{document}